# Two exact derivations of the mass/energy relationship, E=mc$^2$

**Eric Baird**  *(eric_baird@compuserve.com)*

The E=mc$^2$ relationship is not unique to special relativity. Einstein published one exact derivation from special relativity and two approximate derivations that used general extensions to Newtonian mechanics, and an exact derivation is also possible if we use a "first-order" Doppler equation instead of special relativity's "relativistic Doppler" formula. We present two sample derivations based on different non-transverse Doppler relationships, and briefly look at the two diverging systems of physics that result.

## 1. INTRODUCTION

The E=mc$^2$ relationship between mass and energy was first made explicit in a short piece by Einstein ("On the Origin of Inertia" [1]) which was written as a postscript to the famous 1905 "Electrodynamics" paper, [2] and which presented the E=mc$^2$ result as a consequence of the mathematical relationships that had appeared in the earlier piece. W. L. Fadner has also unearthed and discussed a number of contemporary pieces that either came close to deriving E=mc$^2$, or presented similar equations without fully exploring the consequences or claiming the result to be general. [3]

Further pieces by Einstein in the 1940s presented the E=mc$^2$ relationship as a result that could be derived (approximately) from general principles without making specific reference to special relativity, [4][5] and we can also obtain an exact derivation by applying relativistic principles to a different shift formula to the one used by special relativity.

Various authors have reworked the E=mc$^2$ result since. [6][7][8][9][10] What appears to be missing from the literature is a direct comparison between derivations based on these two different shift formulae, and this is what we present here.

## 2. ASSUMPTIONS AND DEPENDENCIES

**Used by both derivations:**

- Momentum of particle,
  mom = mv

- Momentum of light-pressure,
  mom = E / 2c

- Energy of light taken as being proportional to its frequency

- Newton's law of the conservation of momentum

**Used by first derivation only:**

- the "emitter-theory" non-transverse frequency-shift formula,

$$\frac{frequency'}{frequency} = \left( \frac{c - v}{c} \right)$$

**Used by second derivation only:**

- Special relativity's non-transverse frequency-shift formula,

$$\frac{frequency'}{frequency} = \sqrt{\frac{c - v}{c + v}}$$
$$= \left( \frac{c}{c + v} \right) \times \sqrt{1 - \frac{v^2}{c^2}}$$

The variable *v* is taken to be recession velocity throughout.

---

Newton, Optiks 3:1:Qu30: "Are not gross Bodies and Light convertible into one another, and may not Bodies receive much of their Activity from the particles of Light which enter their Composition? For all fix'd Bodies being heated emit Light so long as they continue sufficiently hot, and Light mutually stops in Bodies as often as its Rays strike upon their Parts, as we shew'd above. ... The changing of Bodies into Light, and Light into Bodies, is very conformable to the Course of Nature, which seems delighted with Transmutations. ... And among such various and strange transmutations, why may not Nature change Bodies into Light, and Light into Bodies?"

### 3.    $E=mc^2$, FROM A FIRST-ORDER DOPPLER FREQUENCY-SHIFT FORMULA:

#### Non-radiating particle

A particle travels forwards at $v$ m/s inside a nominally-stationary box.

When the particle hits the rear wall of the box, it gives the box a forward momentum of $mv$. We will set the mass of the particle to be extremely small compared to the mass of the box, so that the change in box velocity caused by the impact is arbitrarily small.

#### Radiating particle

The particle emits a burst of energy $E$ before it hits the box, as two plane waves each with energy $E/2$ in the particle's frame, one traveling "forwards" and one traveling "backwards" along the particle's path. The speed of the particle is not affected by the emission, because the reaction forces of both plane-waves (in the particle's own frame) are equal and opposite.

However, in the frame of the box, the two waves do not have the same energy or momentum. The frequency of the forward-aimed wave is blueshifted by

$$freq' \, / \, freq = (c-(-v)) \, / \, c$$

, the frequency of the rearward-aimed wave is redshifted by

$$freq' \, / \, freq = (c-v) \, / \, c$$

, and the energies of the two waves as they impact on the box are altered by the same amount.

If the individual momentum of each wave is given by $E/2c$, the combined forward momentum of the two waves striking the opposite sides of the box, is calculated as:

$$E \, [ \, (c+v) - (c-v) \, ] \, / \, 2c^2$$

$$= Ev/c^2$$

, so (in the box frame) the two pulses of light emitted by the particle carry $Ev/c^2$ of total forward momentum.

If the particle and its emitted light have the same total overall momentum before and after emission takes place, then after emission the forward momentum of the particle must be reduced by the same amount, $Ev/c^2$.

Since our "spent" particle has reduced momentum but an unchanged velocity, we can apply

$$mass = momentum/velocity$$

to calculate a reduced value for the mass of the particle after emission.

$$mass_1 - mass_2 = (mom_1 - mom_2) \, / \, v$$

$$mass_{LOST} = (Ev/c^2) \, / \, v$$

$$mass_{LOST} = E/c^2$$

In order to lose an amount of inertial mass $m$, a particle with a fixed velocity has to lose an amount of energy (measured in its own frame) equal to

$$E = mc^2$$

---

If we want to try this derivation for pairs of plane-waves aligned at angles *other* than 0° and 180°, we need to know the exact aberration angle-changes associated with our first-order Doppler equation. This derivation will be attempted in a future paper.

### 4.    E=mc², FROM SPECIAL RELATIVITY'S FREQUENCY-SHIFT FORMULA:

#### Non-radiating particle

A particle travels forwards at $v$ $m/s$ inside a nominally-stationary box.

When the particle hits the rear wall of the box, it gives the box a forward momentum of $mv$. We will set the mass of the particle to be extremely small compared to the mass of the box, so that the change in box velocity caused by the impact is arbitrarily small.

#### Radiating particle

The particle emits a burst of energy $E$ before it hits the box, as two plane waves each with energy $E/2$ in the particle's frame, one traveling "forwards" and one traveling "backwards" along the particle's path. The speed of the particle is not affected by the emission, because the reaction forces of both plane-waves (in the particle's own frame) are equal and opposite.

However, in the frame of the box, the two waves do not have the same energy or momentum.
The frequency of the forward-aimed wave is blueshifted by

$$freq'/freq = c \; / \; (c+(-v)) \; \times \; \sqrt{(1 - v^2/c^2)}$$

, the frequency of the rearward-aimed wave is redshifted by

$$freq'/freq = c \; / \; (c+v) \; \times \; \sqrt{(1 - v^2/c^2)}$$

, and the energies of the two waves as they strike the box are altered by the same amount.
If the individual momentum of each wave is given by $E \; / \; 2c$, the combined forward momentum of the two waves striking the opposite sides of the box, is calculated as:

$$E/2c \; \times \; [ \; c/(c-v) - c/(c+v) \; ] \; \times \; \sqrt{(1 - v^2/c^2)}$$
$$= \; [ \; Ev/c^2 \; / \; (1 - v^2/c^2) \; ] \; \times \; \sqrt{(1 - v^2/c^2)}$$
$$= \; Ev/c^2 \; / \; \sqrt{(1 - v^2/c^2)}$$

, so (in the box frame) the two pulses of light emitted by the particle carry $Ev/c^2 \; / \sqrt{(1 - v^2/c^2)}$ of momentum, and if total momentum is conserved, the act of emission must reduce the particle's momentum by this amount.

Since our "spent" particle has reduced momentum but an unchanged velocity, we can apply

$$mass = momentum/velocity$$

to calculate a reduced value for the mass of the particle after emission.

$$mass_1 - mass_2 = (mom_1 - mom_2) \; / \; v$$
$$mass_{LOST} = (Ev/c^2)/v \; / \; \sqrt{(1 - v^2/c^2)}$$
$$mass_{LOST} = E/c^2 \; / \; \sqrt{(1 - v^2/c^2)}$$

We can now rearrange this so that the "$E$" term is on the left. In order to lose an amount of inertial mass $m$, a particle with a fixed velocity $v$ has to lose an amount of energy equal to

$$E = mc^2 \; / \; \sqrt{(1 - v^2/c^2)}$$

Since the total (Lorentz-dilated) mass of this particle under SR [11] is

$$m_{TOTAL} = m_{REST} \; / \; \sqrt{(1 - v^2/c^2)}$$

, we can get rid of the Lorentz term by either:

- making $v=0$ and $m=rest \; mass$, (so that $v^2/c^2=0$, and the Lorentz term disappears) or
- making $v>0$ and using $m$ as Lorentz-dilated mass (in which3 case the two Lorentz terms cancel).

Either way, we then get the more general relationship

$$E = mc^2$$

---

This is essentially Einstein's 1905 derivation, for $f$=0. The concept of Lorentz mass-dilation is now considered by some people to be "old-fashioned", but is still arguably valid for a "historical" derivation.

## 5.    COMPARISON OF THE TWO DERIVATIONS

Although both derivations produce the same nominal result, the resulting physics is different in each case.

### A    First derivation (non-standard)

With the first-order Doppler formula used in Section 3, the wavelength of the forward-emitted wave does not become infinite in the lab frame until the forward velocity of the particle is also infinite. This inability of the particle to catch up and overtake its own wavefront even at nominally "superluminal" velocities means that a model based around this derivation does not use a single fixed speed of light – the speed of a signal leaving the "moving" particle's nose must be different to the speed of a signal emitted by a similar particle that is stationary in the laboratory frame.

If we take the root product of the object's apparent ageing rates when viewed at 0° and 180° to its path, we get an averaged rate of apparent timeflow for the object of $\sqrt{(1 - v^2/c^2)}$, which is the counterpart of the Lorentz time-dilation effect under special relativity.

Emitter-theory and "dragged-light" models should show this behaviour. In contrast to special relativity, these variable-lightspeed models allow speeds up to and beyond $c_{BACKGROUND}$, with the "imaginary" Lorentz result for $v > c_{BACKGROUND}$ being a consequence of the breakdown of direct observations of a superluminally-receding object.

Emitter-theories based on flat spacetime are known not to work.  [12][13][14]

### B    Second derivation (special relativity)

With the so-called "relativistic Doppler" equation (second derivation) the energy of the forward-emitted wave becomes infinite in the background frame as v tends to c, and the requirement that no more energy can be taken out of the experiment than was put in dictates that the amount of energy required to give our particle a velocity v also tends towards infinity as v tends to c.

This still leaves the question of whether the "c" in question refers to $c_{BACKGROUND}$ or to $c_{PARTICLE}$ – these can have different values in a dragged-light model, where $v = c_{PARTICLE}$ represents an unattainable upper limit, with the particle and its forward-emitted light only having the same nominal speed when both have an infinite velocity in the lab frame.

Under special relativity, these issues are avoided by declaring that spacetime is wholly undistorted by relative velocities between physical particles ("spacetime is flat"). The forward-emitted wave traveling at $c_{PARTICLE}$ is then also traveling at $c_{BACKGROUND}$, and the upper limit to the particle's speed has a definite finite velocity in the laboratory frame. There is again a calibrated Lorentz reduction in the apparent rate of timeflow for a moving object, with the rate becoming "imaginary" for $v > c$, but the infinite "energy barrier" at $v = c$ arguably makes this a moot point.

## 6.    CONCLUSIONS

Because Einstein's "inertia" paper was published shortly after the electrodynamics paper, and presented the E=mc$^2$ relationship as a consequence of the equations in the earlier piece, it is perhaps natural to suppose that the E=mc$^2$ relationship would not have been discovered without special relativity.

However, E=mc$^2$ is also a consequence of the first-order Doppler equations associated with Newtonian ballistic-photon models, and, given Newton's published hypothesis about the interconvertibility of matter and light, [15] the significance of the relationship could have been recognised much earlier. Fadner's research [3] shows that some contemporary papers did include near-derivations, or derivations that were presented as limited special cases.

If we consider the three most fundamental non-transverse Doppler equations (the two basic first-order equations and special relativity's intermediate "relativistic Doppler" equation), only one of the three has *not* been shown to give us an exact derivation of E=mc$^2$ – and of the two derivations given, special relativity's is not the shortest. Although special relativity deserves credit for providing a credible theoretical platform that allowed Einstein to derive and publish the result, a verification of E=mc$^2$ is not necessarily a verification of special relativity.